\documentclass[onecolumn]{article}

\usepackage{cite}
\usepackage[title]{appendix}

\usepackage{graphicx, amssymb, amsmath}
\usepackage[T1]{fontenc}   
\usepackage[utf8]{inputenc}
\usepackage{lmodern}
\usepackage[unicode]{hyperref}
\hypersetup{pdfnewwindow}

\usepackage{isomath}
\usepackage{xr}

\let\vec\vectorsym
\let\tens\tensorsym

\newcommand{\uvec}[1]{{\vec{u}_{#1}}}
\newcommand{\intr}[1]{\bar{#1}}
\newcommand{\xma}[1]{{\tens{#1}_{\scriptscriptstyle\times}}}

\newcommand{\ff}{{\it ff}}

\newcommand{\diag}{{\rm diag}}
\newcommand{\com}{{\scriptscriptstyle{\mathbb{C}}}}

\newcommand{\eff}{{\it eff}}
\renewcommand\Re{\operatorname{Re}}
\renewcommand\Im{\operatorname{Im}}

\let\originalleft\left
\let\originalright\right
\renewcommand{\left}{\mathopen{}\mathclose\bgroup\originalleft}
\renewcommand{\right}{\aftergroup\egroup\originalright}

\newcommand{\rmi}{\mathrm{i}}

\newcommand{\eref}[1]{({\ref{#1}})}

\begin{document}

{\title{Analysis of the measurements of anisotropic a.c. vortex resistivity in tilted magnetic fields.}
}

\author{Nicola~Pompeo, Enrico~Silva%
\thanks{N. Pompeo and E. Silva are with the Dipartimento di Ingegneria, Universit\`a Roma Tre, Via V. Volterra 62, 00146 Roma, Italy.}
}

\providecommand{\keywords}[1]{\textbf{\textit{Index terms---}} #1}
\maketitle

\begin{abstract}
Measurements of the high--frequency complex resistivity in superconductors are a tool often used to obtain the vortex parameters, such as the vortex viscosity, the pinning constant and the depinning frequency. In anisotropic superconductors,
the extraction of these quantities from the measurements faces new difficulties due to the tensor nature of the electromagnetic problem. The problem is specifically intricate when the magnetic field is tilted with respect to the crystallographic axes. Partial solutions exist in the free--flux--flow (no pinning) and Campbell (pinning dominated) regimes. In this paper we develop a full tensor model for the vortex motion complex resistivity, including flux--flow, pinning, and creep. We give explicit expressions for the tensors involved. We obtain that, despite the complexity of the physics, some parameters remain scalar in nature. We show that under specific circumstances the directly measured quantities do not reflect the true vortex parameters, and we give procedures to derive the true vortex parameters from measurements taken with arbitrary field orientations. Finally, we discuss the applicability of the angular scaling properties to the measured and transformed vortex parameters and we exploit these properties as a tool to unveil the existence of directional pinning.
\end{abstract}


\maketitle

\section{Introduction}
Material anisotropy, which characterizes many superconductors of wide interest  \cite{Aswathy2010, Buzea2001, PooleJr2007}, has a profound impact
on the electrical transport properties and, in particular,
on the vortex dynamics and on the related pinning phenomena. 
The tailoring of superconductive materials in order to reduce their 
apparent
({\em effective}) anisotropy and the vortex-induced dissipation, avoiding detrimental effects on other superconducting properties, is an actively pursued research \cite{Gutierrez2007a, Obradors2014, Rizzo2016}.

High frequency measurements of the electric transport properties in the mixed state provide a powerful tool to investigate
the vortex motion dissipation, related to the vortex viscosity (or viscous drag), and vortex pinning, related to the Labusch parameter and the creep factor \cite{Golosovsky1996}.

When anisotropic properties are investigated, measurements are performed by tilting the applied static magnetic field $H$ with respect to the anisotropy axes and the current direction. Then, a complicated tensor model emerges, and the analysis of the data is not straightforward.

Previous works addressed separately the flux--flow regime (pinning neglected) and the pinned regime.

The d.c. flux flow regime has been thoroughly investigated, so that the flux flow
resistivity tensor and the related vortex viscosity tensor are well characterized \cite{ Hao1996a,Ivlev1991,Genkin1989,Hao1991,Bespalov2012}. 
In the a.c. low frequency regime, pinning is dominant. Despite the many existing theories \cite{Blatter1994,Larkin1979a,Bennemann2008},
this regime has been addressed only partially\cite{Klupsch1992, Shklovskij2006, Dobrovolskiy2016}.

In \cite{Pompeo2015}
a full description of the complex resistivity in the low frequency, a.c. linear regime where pinning is dominant (Campbell regime) was developed. The model incorporated the material uniaxial anisotropy and point pinning in the vortex a.c. resistivity tensor with a magnetic field applied at an arbitrary orientation.

In this paper
we address another important issue: we develop a model for the tensor complex resistivity in the high-frequency range (typically rf and microwaves), where both dissipation (flux--flow) and pinning give comparable contributions and must be considered at the same time. We also include the effect of flux-creep, that can give a detectable effect \cite{Brandt1991,Coffey1991a, Silva2011, Pompeo2007a, Pompeo2012, Song2009}.

The paper is organized as follows.
In Section \ref{sec:scalarmodel} we recall the well--known scalar model for the a.c. vortex motion resistivity. Section \ref{sec:acforceequation} concisely reports the formulation of the tensor model in the Campbell regime \cite{Pompeo2015} and introduces the tensor notation. The full anisotropic tensor model, which includes flux--flow, pinning and flux creep, is developed and described in Section \ref{sec:rhovmtensor}.
In Section \ref{sec:rhovm_examples} we 
connect our model to the experiments,
providing examples of experimental data analysis.
In Section \ref{sec:summary} we summarize and present possible extensions of this work.
In the Appendixes, we provide (A) a short application of the formalism to the commonly used setup with straight currents and (B) a summary list of the most useful expressions for the analysis of vortex resistivities.
\section{A.c. vortex motion resistivity tensor: the framework}
\label{sec:highfrequency}

In this Section we present the framework for the development of the full tensor model for the complex vortex resistivity ${\rho}_v$. To do so, in Section \ref{sec:scalarmodel} we first shortly recall the scalar models for high--frequency vortex dynamics, useful for the introduction of the vortex parameters.
The complications of the tensor model for uniaxial superconductors with a field applied in an arbitrary orientation are introduced in Section \ref{sec:acforceequation}, limited to the pure (flux flow or Campbell) regimes.
We devote particular care to the identification of the experimentally obtainable parameters and to the comparison with theoretical quantities.
\subsection{Short review of scalar models}
\label{sec:scalarmodel}

Vortices nucleated from the magnetic induction ${B}$ 
and subjected to a current field
${J}$ experience a force (per unit length), so that fluxon motion arises. 
The scalar force equation for the vortex velocity $v$ in an isotropic superconductor with isotropic (point) pinning and $B$ perpendicular to $J$ is, in the harmonic regime $e^{\rmi\omega t}$  \cite{Gittleman1966,Golosovsky1996}:
\begin{equation}
\label{eq:forceIso}
\eta v+\frac{k_p}{\rmi\omega}v=\Phi_0 J+F_{therm}
\end{equation}
where
$\eta$ is the drag coefficient (vortex viscosity) and $k_p$ is the pinning constant (Labusch parameter). Together, they define the (de)pinning angular frequency $\omega_p=k_p/\eta$ and the depinning characteristic time $\tau_p=\omega_p^{-1}$.
$F_{therm}$ is a stochastic thermal force causing thermal depinning (vortex creep)
with an activation characteristic time $\tau_{th}$ dependent on the activation energy $U$.
Different approaches \cite{Brandt1991,Coffey1991a, Coffey1992}, with different ranges of applicability \cite{Pompeo2008}, have been proposed to take into account creep effects.
It is possible to develop a generalized model, independent on the specific choice for $\tau_{th}$ \cite{Pompeo2008}. One then obtains the vortex motion complex resistivity as:
\begin{equation}
\label{eq:rhoBiso}
\rho_{v}=\frac{\Phi_0 B}{\eta}\frac{\chi+\rmi\frac{\omega}{\omega_0}}{1+\rmi\frac{\omega}{\omega_0}}=\frac{\Phi_0 B}{\eta_\com}
\end{equation}
where the latter equality defines a {\em complex viscosity} $\eta_\com$. In the generalized Equation \eqref{eq:rhoBiso}, $\chi(U/K_BT)$ is a creep factor, and the characteristic frequency $\omega_0\rightarrow \omega_p$ for $\chi\rightarrow 0$.

In the specific model developed by E. H. Brandt \cite{Brandt1991}, which will prove useful later on, thermal depinning is described in terms of the relaxation of the pinning constant $k_p(t)=k_p e^{-t/\tau_{th}}$, and one has:
\begin{equation}
\label{eq:etac0}
\eta_\com=\eta\left(1-\rmi\frac{\omega_p}{\omega}\frac{1}{1-\frac{\rmi}{\omega\tau_{th}}}\right)
\end{equation}
for the complex viscosity, with the characteristic angular frequency $\omega_0=\tau_{th}^{-1}+\tau_p^{-1}$ and the creep factor $\chi=1/{(1+e^{U/K_BT})}$.
For $U\rightarrow\infty$ the creep is negligible, $\chi\rightarrow0$ and $\omega_0\rightarrow\omega_p$, so that \eqref{eq:rhoBiso} becomes: 
\begin{equation}
\label{eq:rhoGRiso}
\rho_{v}=\frac{\Phi_0 B}{\eta}\frac{1}{1-\rmi\frac{\omega_p}{\omega}}=\left(\rho_\ff^{-1}-\rmi\rho_C^{-1}\right)^{-1}
\end{equation}
where $\rho_{\ff}=\Phi_0B/\eta$ and $\rho_C=\omega\Phi_0B/k_p$ are the flux--flow and the Campbell resistivity, respectively.
This limit corresponds to the Gittleman\textendash{}Rosenblum (GR) model  \cite{Gittleman1966}. From \eref{eq:rhoGRiso} it can be seen that $\omega_p$ marks the transition between a ``low frequency'' and a ``high frequency'' regime: for $\omega\ll\omega_p$ the pinning force dominates over the viscous drag, yielding $\rho_{v}\rightarrow\rmi\rho_C$, while for $\omega\gg\omega_p$, 
$\rho_{v}\rightarrow\rho_\ff$ corresponding to a purely dissipative flux flow regime analogous to the one observable in d.c. with no pinning.

Before concluding this review, it is worth recalling the often used  dimensionless ratio $r$ \cite{Halbritter1995,Tsuchiya2001a,Velichko2002,Pompeo2009,Song2009,Silva2015,Torokhtii2016}:
\begin{equation}
\label{eq:r}
r=\frac{\Im(\rho_{v})}{\Re(\rho_{v})}
\end{equation}
which, if creep is negligible (GR limit), yields:
\begin{equation}
\label{eq:rGR}
r=\frac{\rho_C}{\rho_{\ff}}=\frac{\omega_p}{\omega}
\end{equation}
The $r$ parameter can be directly computed from the complex resistivity $\rho_v$ and it is unaffected by any systematic scale factor in the experiments. Physically, it allows to easily evaluate whether the vortex dynamics is in the pinning ($r\gg1$) or flux flow ($r\ll1$) dominated regime.

\subsection{The anisotropic model framework}
\label{sec:acforceequation}

\begin{figure}[ht]
\centerline{\includegraphics[width=4cm]{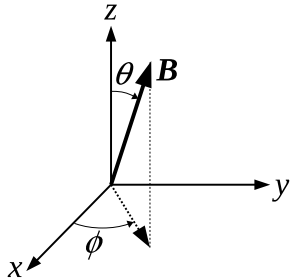}}
  \caption{ Frame of reference. The magnetic induction field $\vec{B}$ is depicted, applied along a generic direction identified by the polar and azimuthal angles $\theta, \phi$.}
\label{fig:frame}
\end{figure}

We treat the case of a superconductor with uniaxial anisotropy along the $c$--axis. We choose a reference frame such as $x\equiv a$, $y\equiv b$ and $z\equiv c$. Vectors are denoted as $\vec{A}=\uvec{A}A$, where $\uvec{A}$ and $A$ are the unit vector and the modulus, respectively. A tensor/matrix is denoted as $\tens{A}$. Vector
orientation
can be identified by means of the polar $\theta$ and azimuthal $\phi$ angles. Figure \ref{fig:frame} illustrates the reference frame together with the magnetic induction field vector $\vec{B}=B\uvec{B}$ as an example.

We briefly recall the notation and the results of \cite{Pompeo2015}. In the London limit, the only source of material anisotropy is given by
the phenomenological electronic mass tensor\cite{Klemm1980,Kogan1981}: $m_{ab}\tens{M}=\diag(m_{ab},m_{ab},m_c)$, where $m_{ab}$ and $m_c$ are the in-plane and out-of-plane effective mass of the charge carriers, respectively.
Other possible sources of anisotropy, such as the scattering time of the normal carriers, are neglected, 
as well as ``exotic'' anisotropic properties such as the nonreciprocal magnetochiral anisotropy in non-centrosymmetric superconductors \cite{Sato2017, Wakatsuki2017}.

Vortices moving with a given velocity $\vec{v}$ ($\perp \vec{B}$) induce (Faraday's law) an electric field $\vec{E}=\vec{B}\times\vec{v}$ which determines
a \emph{response} transport current density $\vec{J}_T=\tens{\intr{\sigma}}\vec{E}$, where $\tens{\intr{\sigma}}$ represents the ``intrinsic'', material dependent conductivity
\cite{Hao1993a,Hao1996a}.
However, in most experiments, vortices are set in motion by the applied current density $\vec{J}$ by means of the Lorentz force. In this case
the response of the superconducting system to the applied current is the electric field induced by vortex motion. Hence $\vec{E}=\tens\rho\vec{J}$, where in this case the system ``transfer function'' is given by the resistivity tensor $\tens\rho$.
The experimentally measured $\tens\rho$ and the material property $\tens{\intr\rho}=\left(\tens{\intr\sigma}\right)^{-1}$ are related by (we neglect Hall effect) \cite{Pompeo2015}:
\begin{equation}
\label{eq:rhomain}
\tens{\rho}=-\xma{B}\left(\frac{\left|\tens{\intr{\rho}}\right|\tens{\intr\rho}^{-1}}{\left(\tens{\intr{\rho}}\uvec{B}\right)\cdot\uvec{B}}\right)\xma{B}
\end{equation}
where $\xma{B}=\tens{\epsilon}_{ijk}\uvec{B}$, and $\tens{\epsilon}_{ijk}$ is the permutation (Levi-Civita) tensor \cite{Itskov2009}.
The difference between $\tens\rho$ and $\tens{\intr\rho}$ stems from the Lorentz--Faraday origin of the force, and thus depends on the $\vec{B}$ orientation.
Therefore the experimentally measured $\tens\rho$ exhibits a dependence on the $\vec{B}$ orientation of double origin: from the material ``intrinsic'' property $\tens{\intr{\rho}}$ and from the ``extrinsic'' Lorentz-Faraday terms.

An important remark is that
the actual expression of the material dependent $\tens{\intr\rho}$ depends
on the force equation, and then
on the particular vortex motion regime considered. 
This feature is common to other vortex parameters, as we show in the following. 

In the pure flux flow regime, the force equation (in tensor form) reads:
\begin{equation}
\label{eq:force_eta}
\tens{\eta}\vec{v}=\Phi_0 \vec{J}\times\uvec{B}
\end{equation}
with
\begin{subequations}
\label{eq:etatensor}
\begin{align}
\tens{\eta}&=-\xma{B}\tens{\intr\eta}\xma{B}\\
\tens{\intr\eta}&=\Phi_0 B \tens{\intr{\sigma}}_\ff
\end{align}
\end{subequations}
where the full theoretical expression of $\tens{\intr\sigma}_\ff=\left(\tens{\intr\rho}_{\ff}\right)^{-1}$ (the subscript ``$\ff$'' refers to flux flow regime related quantities) can be computed within the Time Dependent Ginzburg-Landau theory \cite{Hao1993a, Hao1996a,Ivlev1991,Genkin1989,Hao1991}. \\
In the a.c., low frequency Campbell regime\cite{Campbell1969a}
the elastic recall by pinning centers dominates the vortex motion. Assuming that no other preferential directions are introduced by the pinning centers, that is that pinning originates from point pins only, making reference to \eqref{eq:forceIso} and taking into account the vector form of the force equation one gets:
\begin{equation}
\label{eq:force_kp}
\frac{1}{\rmi\omega}\tens{k}_p\vec{v}=\Phi_0 \vec{J}\times\uvec{B}.
\end{equation}
Expressions analogous to \eqref{eq:etatensor} can be obtained \cite{Pompeo2015}:
\begin{subequations}
\label{eq:kptensor}
\begin{align}
\tens{k}_p&=-\xma{B}\tens{\intr k}_p\xma{B} \\
\tens{\intr k}_p&=\omega\Phi_0B\tens{\intr\sigma}_{C}
\end{align}
\end{subequations}
where $\tens{\intr\sigma}_C=\left(\tens{\intr\rho}_{C}\right)^{-1}$ are the Campbell conductivity and resistivity intrinsic tensors.
The field and temperature dependent explicit expressions for the above pinning tensors can be worked out within the theory for collective point pinning in anisotropic superconductors\cite{Blatter1994,Larkin1979a}. 

It is possible to further elucidate the angular dependence of the intrinsic quantities even without explicitly writing down the full theoretical expressions. Indeed, the largely used Blatter-Geshkenbein-Larkin (BGL) scaling theory \cite{Blatter1992,Vinokur1993a, Geshkenbein1994} states that $\tens{\intr\rho}_{\ff}$, $\tens{\intr\eta}$, $\tens{\intr\rho}_{C}$, $\tens{\intr k}_p$ obey the following scaling laws in the point pinning regime:

\begin{subequations}
\label{eq:etatensor_scaling}
\begin{align}
\label{eq:etatensor_scalinga}
\tens{\intr\rho}_\ff(B,\theta)&=\intr\rho_{\ff,11}(B,\theta)\tens{M}={\intr\rho}_{\ff,11}(B\epsilon(\theta),0)\tens{M} \\
\label{eq:etatensor_Bdep}
\tens{\intr\eta}(B,\theta)&=\intr \eta_{11}(B,\theta)\tens{M}^{-1}=\frac{\intr\eta_{11}(B\epsilon(\theta),0)}{\epsilon(\theta)}\tens{M}^{-1}
\end{align}
\end{subequations}
\begin{subequations}
\label{eq:kptensor_scaling}
\begin{align}
  \tens{\intr\rho}_{C}(B,\theta)&=\intr\rho_{C,11}(B,\theta)\tens{M}={\intr\rho}_{C,11}(B\epsilon(\theta),0)\tens{M} \\
\label{eq:kptensor_Bdep}
  \tens{\intr k}_p(B,\theta)&=\intr k_{p,11}(B,\theta)\tens{M}^{-1}=\frac{\intr k_{p,11}(B\epsilon(\theta),0)}{\epsilon(\theta)}\tens{M}^{-1}
\end{align}
\end{subequations}
where $\gamma^2=m_c/m_{ab}$ is the mass anisotropy factor and 
\begin{equation}
\label{eq:epsilon}
\epsilon(\theta)=(\cos^2\theta+\gamma^{-2}\sin^2\theta)^{1/2}
\end{equation}
is the angular-dependent anisotropy parameter. Hence, the combination of \eqref{eq:rhomain} with either \eqref{eq:etatensor_scaling} or \eqref{eq:kptensor_scaling} allows to fully describe the resistivity angular dependence for $\vec{B}$ and $\vec{J}$ with arbitrary orientations, provided that the pure flux flow or Campbell regime are considered. We now extend the model to more general regimes.

\section{The full a.c. vortex motion anisotropic model}
\label{sec:rhovmtensor}

We here extend the model to the case of simultaneous relevance of both flux--flow and elastic recall regimes. For the sake of clarity, we neglect temporarily the thermal effects (flux creep), that we reintroduce later on.
The force equation in anisotropic superconductors including both viscous drag and pinning is:
\begin{equation}
\label{eq:force_vm1}
\tens{\eta}\vec{v}+\frac{1}{\rmi\omega}\tens{k}_p\vec{v}=\Phi_0 \vec{J}\times\uvec{B}
\end{equation}
By defining a complex viscosity tensor as:
\begin{equation}
\label{eq:etac1}
\tens{\eta}_{\com}=\tens{\eta}-\rmi\frac{\tens{k}_p}{\omega}
\end{equation}
the above equation can be recast as:
\begin{equation}
\label{eq:force_vm2}
\tens{\eta}_\com\vec{v}=\Phi_0 \vec{J}\times\uvec{B}
\end{equation}
The force equation \eref{eq:force_vm2} is formally equivalent to force equations written for the flux flow \eref{eq:force_eta} and Campbell regimes  \eref{eq:force_kp}.
Hence, the previous results can be straightforwardly extended to write, as a first result:
\begin{equation}
\label{eq:etac2}
\tens{\eta}_{\com}=-\xma{B}\tens{\intr\eta}_{\com}\xma{B}
\end{equation}
Equations \eref{eq:etac1} and \eref{eq:etac2} allow to write down:
\begin{equation}
\label{eq:etaci1}
\tens{\intr\eta}_{\com}=\tens{\intr\eta}-\rmi\frac{\tens{\intr k}_p}{\omega}\\
\end{equation}
Substituting \eref{eq:etatensor_Bdep} and \eref{eq:kptensor_Bdep}  into \eref{eq:etaci1} yields:
\begin{align}
\label{eq:etaci2} 
\nonumber
\!\tens{\intr\eta}_{\com}\!(B, \theta)\!&=\!\left(\!\intr\eta_{11}(B,\theta)\!-\!\rmi \frac{\intr k_{p,11}(B,\theta))}{\omega}\!\right)\!\tens{M}^{-1}\!=\\ 
&=\left(1-\rmi\frac{\omega_p(B,\theta)}{\omega}\right)\tens{\intr\eta}(B,\theta)
\end{align}
It is evident that $\tens{\intr\eta}_{\com}$ inherits the angular dependencies and anisotropic properties from $\tens{\intr\eta}$ and $\tens{\intr k}_p$, represented by $\tens{M}$ and by the scaling law. 
Moreover, by computing \eref{eq:etaci2}, the first important result of this paper, particularly relevant in the analysis of the experiment, emerges. Namely:
\begin{equation}
\label{eq:omegap}
\frac{{\intr k}_{p,ii}(B,\theta)}{\intr\eta_{ii}(B,\theta)}\!=\!\omega_p(B,\theta)=\omega_p(B\epsilon(\theta),0)
\end{equation}
which holds for $i=1...3$ (all the principal axes directions).

That is: 
contrary to the viscosity or the pinning constant, the pinning frequency is a scalar
quantity, and it is affected by the direction of the magnetic field with respect to the crystal axes only through the rescaled field $B\epsilon(\theta)$.
This fact implies that, whichever orientation is set for $\vec{B}$ and $\vec{J}$, the vortex system will always have the same pinning frequency at fixed $B/B_{c2}(\theta)$.

The property exemplified by \eqref{eq:omegap}
suggests a straightforward method to check whether directional defects influence vortex motion.
In a plot of
$\omega_p$ vs the scaled field $B/B_{c2}(\theta)$, any deviation from a scaling curve
would be a fingerprint for
some directional effects other than the material anisotropy.
In fact, should extended pins be present,
\eref{eq:kptensor_Bdep} would not hold.
This method is a second outcome of this work.
It is particularly useful to the analysis of the experiments once it is recognized that, within ample margins, a nonzero creep factor does not change much the derived values of the pinning frequency \cite{Pompeo2008}. Then, using the GR model to obtain $\omega_p$ yields small uncertainties, and since in the GR model the experimental parameter $r=\omega_p/\omega$ one has a direct, purely experimental quantity to reveal directional effects other than the mass anisotropy.

A third result is that, with
point pinning, the vortex complex resistivity tensor retains the same anisotropic behavior as the flux--flow or Campbell resistivity tensors.
In fact, the conductivity and resistivity tensors are:
\begin{equation}
\label{eq:rho_vi1}
\tens{\intr\rho}_v=\left(\tens{\intr\sigma}_{v}\right)^{-1}=\Phi_0B \left(\tens{\intr\eta}_{\com}\right)^{-1}
\end{equation}
Using \eref{eq:etaci2}, one derives the explicit expression for $\tens{\intr\rho}_v$:
\begin{subequations}
\label{eq:rho_vi2}
\begin{align}
\label{eq:rho_vi2a}
\tens{\intr\rho}_v(B,\theta)&=\intr\rho_{v,11}(B,\theta)\tens{M}=\intr\rho_{v,11}(B\epsilon(\theta),0)\tens{M} \\
\label{eq:rho_vi2b}
\intr\rho_{v,11}(B,\theta)&=\intr\rho_{\ff,11}(B,\theta)\frac{1}{1-\rmi\frac{\omega_{p}(B,\theta)}{\omega} }
\end{align}
\end{subequations}
It can be noted that, thanks to the scalar nature of the pinning angular frequency $\omega_p$, the tensors $\tens{\intr\rho}_v(B,\theta)$ and $\tens{\intr\rho}_{\ff}(B,\theta)$ differ by a \emph{scalar} quantity, $(1-\rmi\omega_{p}(B,\theta)/\omega)$.
Thus, similarly to $\tens{\intr\eta}_{\com}$,  $\tens{\intr\rho}_v$ retains the same anisotropy of the flux flow and pinning tensors, given by the mass anisotropy tensor $\tens{M}$ alone. Moreover, it satisfies the same angular scaling law, shown in the last equality, as $\tens{\intr\rho}_\ff$ and $\tens{\intr\rho}_C$.

It is important to remember that $\tens{\intr\rho}_v$ of \eref{eq:rho_vi2} is not directly measured, whereas the actually measured tensor is computed through \eqref{eq:rhomain}:
\begin{equation}
\label{eq:rho_v}
\tens{\rho}_{v}(B,\theta,\phi)=\intr\rho_{v,11}(B,\theta)\left[\frac{\tens{M}_B(\theta,\phi)}{\epsilon^2(\theta)}\right]
\end{equation}
where $\tens{M}_B(\theta,\phi)=-\xma{B}\tens{M}^{-1}\xma{B}$ has been introduced for the sake of compactness.
From \eref{eq:rho_v} it can be noted that, contrary to $\tens{\intr\rho}_v$, the measured tensor $\tens{\rho}_{v}$ does not obey the angular scaling law, since it incorporates additional angular dependencies through $\tens{M}_B$
(we recall that such additional angular dependencies derive, ultimately, from the vector form of the Lorentz force).

Now we include the effects of flux creep. The pinning energy $U$ depends only on the magnetic field magnitude and direction and not on the direction of vortex motion. Moreover, it obeys the usual scaling law with a constant scaling factor $\gamma^{-1}$ \cite{Blatter1992}. Therefore the thermal depinning time $\tau_{th}$ (Section \ref{sec:scalarmodel}), being computed from scalar quantities $U$ and $\omega_p$, is a scalar. The pinning constant tensor \eref{eq:kptensor_Bdep} can thus be modified to include creep as:
\begin{equation}
\label{eq:kpicreep}
\tens{\intr k}_p\!(B,\theta)\!=\!\intr k_{p,11}(B,\theta)\left(\!1\!-\frac{\rmi}{\omega\tau_{th}\!(B,\theta)}\right)\!\tens{M}^{-1}
\end{equation}
Consequently, the scalar vortex motion resistivity with flux creep \eref{eq:rhoBiso} can be generalized to the anisotropic case
in the same way as in the creep-free case. One obtains the same \eref{eq:rho_vi2a} and \eref{eq:rho_v} but with a different expression for the intrinsic vortex motion resistivity:
\begin{equation}
\label{eq:rho_vi3}
\intr\rho_{v,11}(B,\theta)=\intr\rho_\ff\frac{\chi+\rmi\frac{\omega}{\omega_{0}}}{1+\rmi\frac{\omega}{\omega_{0}}}
\end{equation}
where $\omega_0$ and $\chi$ are scalar as in the isotropic case.
Hence, the tensors $\tens{\intr\rho}_v(B,\theta)$ and $\tens{\intr\rho}_{\ff}(B,\theta)$ differ by a scalar factor as in the zero creep case.
This property will prove important in the interpretation of the experiments.

It is worth stressing that the choice of the pinning constant relaxation as a model for flux creep \cite{Brandt1991} is not a limiting factor to the results obtained up to now: in fact, any pure thermal creep process (independent on the angle between 
$\vec{B}$ and $\vec{J}$), possibly with a different definition of $\chi$ and $\omega_0$ \cite{Pompeo2008}, would yield the same results.

\section{Application to experiments: the measured complex vortex resistivity in common setups}
\label{sec:rhovm_examples}

In this Section we consider explicitly some typical experimental configurations and we derive specific expressions relating measured and material intrinsic quantities.
We assume to be deep in the mixed state, but far from pair-breaking effect, so that the contribution of the coupled superconducting and normal microwave currents is negligible.
In this case, the change in the response due to the application of the magnetic field is due to the vortex motion: superfluid and quasiparticle contributions give a nearly field--independent contribution that can be easily subtracted out.
Moreover, we assume the superconducting material to be a thin film, a sample geometry widely used in microwave experiments \cite{Revenaz1994, Ghosh1997, Powell1998, Banerjee2004, Silva2003c, Janjusevic2006}. In these conditions it can be shown\footnote{
The derivation in a scalar problem has been discussed since longtime \cite{Pompeo2017c, Pompeo2017d}. It can be proven that the same holds also for the anisotropic problem, but the lengthy derivation is outside of the scope of this paper and it will be reported elsewhere.
}
that $\tens{\rho}_v$, being essentially proportional to the surface impedance tensor, can be taken as the actually measured quantity.

\subsection{Rotational symmetric planar currents}
\label{sec:rhovm_examples2}

The frequency range where both flux--flow and Campbell contributions have a similar weight is around the pinning frequency $\omega_p/2\pi$.
For example, in
YBa$_2$Cu$_3$O$_{7-\delta}$ $\omega_p/2\pi=1\div 30$ GHz \cite{Willemsen1995, Golosovsky1996, Belk1997, Powell1998, Tsuchiya2001a, Torokhtii2017}, depending on the purity of the material and on the temperature,
while in low-$T_c$ superconductors it ranges from a few MHz \cite{Gittleman1966} to a few GHz, as measured in thin films \cite{Janjusevic2006,Awad2011,Pompeo2012}.
Thus, the microwave range is the frequency regime of interest. 
The high sensitivity that can be achieved in microwave measurements due to the use of cylindrical resonators \cite{Mazierska2001} made the rotational--symmetric current pattern a very popular choice for measurements in superconducting films
\cite{Tsuchiya2001a, Pompeo2014}. In this case, the microwave modes TE$_{0xy}$ are used (more often, $0xy=011$), and the microwave current flows along circular paths in the isotropic planes.

Another technique, preferred for its broadband capability, is the so--called Corbino disk \cite{Silva2011,Booth1994, Wu1995, Scheffler2005a,   Pompeo2012,Torokhtii2012, Torokhtii2012a, Silva2016}, where a planar sample short--circuits a transmission line, giving rise to radial currents on the isotropic plane.
\begin{figure}[ht]
\centerline{(a)\;\includegraphics[width=0.6\columnwidth]{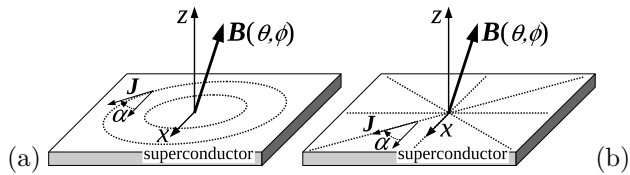}(b)}
  \caption{
  Most common experimental configurations with rotational--symmetric currents. (a): typical configuration with the resonator end--plate--replacement technique, in a TE$_{0xy}$ mode: $\vec{J}$ flows along circular lines, in the isotropic plane. (b): current configuration in the Corbino disk geometry: $\vec{J}$ flows radially, in the isotropic plane.
}
\label{fig:rotational}
\end{figure}

The two mentioned cases are sketched in Figure \ref{fig:rotational}. In both cases
the actually measured \cite{Pompeo2012a, Pompeo2012b, Pompeo2013} resistivity comes out as an angular average \cite{Collin1992} over the current pattern: the result has been presented in \cite{Pompeo2015} for the specific case of the Campbell resistivity tensor $\tens{\rho}_C$,
but the derivations in Section \ref{sec:rhovmtensor} make the consideration valid for the full vortex motion resistivity tensor $\tens{\rho}_v$.
We can then write the expression for the vortex resistivity averaged over a rotational symmetric current distribution:
\begin{subequations}
\label{eq:rhoeff_exp53}
\begin{align}
\label{eq:rhoeff_exp53a}
\rho_{v}^{(\circ)}(B,\theta)&=\intr\rho_{v,11}(B,\theta)f_{L}(\theta)\\
\label{eq:rhoeff_exp53b}
f_{L}(\theta)&=\frac{\frac{1}{2}\gamma^{-2}\sin^2\theta+\cos^2\theta}{\gamma^{-2}\sin^2\theta+\cos^2\theta}
\end{align}
\end{subequations}
It can be seen that the effective, measured vortex motion resistivity consists in the product of two terms: the first is the resistivity $\intr\rho_{v,11}$ only, which in particular obeys the angular scaling law; the second one, denoted in the equation as $f_{L}(\theta)$, is an additional angular dependence which arises from the Faraday and Lorentz actions. Consequently, as already anticipated commenting the whole tensor $\tens{\rho}_v$, the experimentally measured quantity does not obey the scaling law. Therefore care must be taken in separating the intrinsic material property from the contribution given by the setup geometry before proceeding with the physical interpretation of the data.

The most important results coming from \eqref{eq:rhoeff_exp53} is that
the evaluation of the anisotropy from a system where the currents have rotational symmetry is affected by the Lorentz-originating angular contribution, and the apparent anisotropy can be larger \cite{Pompeo2013, Silva2015c} than the intrinsic (mass tensor) anisotropy.

Another commonly used current pattern consists in straight currents. The obvious advantage of probing individual spatial orientations is counterbalanced by the reduced sensitivity that can be achieved with resonators supporting the required electromagnetic field geometry. The full derivation of the relevant expressions, leading to results analogous to those of \eqref{eq:rhoeff_exp53}, is
given in Appendix \ref{sec:rhovm_examples1}.

\subsection{Angle-dependent effective quantities}
It is interesting to note that in a typical GR (no-creep) model analysis for an isotropic superconductor the vortex parameters $\rho_\ff$, $r$ and $k_p$ are extracted from the complex measured $\rho_{v}$ following \eref{eq:rhoGRiso}, yielding:
\begin{subequations}
\label{eq:rhoGR2}
\begin{align}
\label{eq:rhoGR2a}
  \rho_{\ff} & =\Re(\rho_{v}) \left[1+\left(\frac{\Im(\rho_{v})}{\Re(\rho_{v})}\right)^2\right] \\
  \label{eq:rhoGR2b}
  r &=\frac{\Im(\rho_{v})}{\Re(\rho_{v})} \\
  \label{eq:rhoGR2c}
  k_p &=\frac{r}{\rho_\ff}\omega B \Phi_0 = \omega B \Phi_0 \frac{\Im(\rho_{v})}{\Re^2(\rho_{v})+\Im^2(\rho_{v})}
\end{align}
\end{subequations}
On the other hand, when performing measurements on an anisotropic superconductor probed with rotational symmetric current patterns leading to \eref{eq:rhoeff_exp53}, this computation would yield the following {\em effective} quantities (apart from an additional $\phi$-dependence, the same holds for the straight current setup described by \eref{eq:rhoeff_exp4}):
\begin{subequations}
\label{eq:rhoGR3}
\begin{align}
\label{eq:rhoGR3a}
  \rho_{\ff,\eff}(B,\theta) & =\intr\rho_{\ff,11}(B,\theta) f_{L}(\theta) \\
\label{eq:rhoGR3b}
  r_{\eff}(B,\theta) &=r(B,\theta) \\
\label{eq:rhoGR3c}
  k_{p,\eff}(B,\theta) &=\frac{\intr k_{p,11}(B,\theta)}{f_{L}(\theta)}
\end{align}
\end{subequations}
It can be seen that the parameter $r=\omega_p/\omega$ is directly obtained from the measured quantities: this is an interesting result, which allows a direct evaluation of the material anisotropy of the system without the need to deal with Lorentz-dependent contribution $f_{L}(\theta)$. On the other hand, both $\rho_{\ff,\eff}$ and $k_{p,\eff}$ show an additional angular dependence through $f_{L}(\theta)$. Therefore, in the analysis of angular data care must be devoted in correctly extracting the intrinsic quantities, as opposed to effective quantities.
This requires to evaluate the $f_{L}(\theta)$ function, which in turn requires the knowledge of the anisotropy factor $\gamma$.

Further information can be gained through a discussion of the quantities \eref{eq:rhoGR3} in terms of angular scaling. First, we note that the experimental $r_{\eff}(B,\theta)$ coincides with the intrinsic material property, independent on the field--current--crystal angle ($f_L(\theta)$). Thus, in absence of directional effects other than the mass anisotropy, it should scale as $r_{\eff}(B,\theta)=r_{\eff}(B\epsilon(\theta))$. If it does not, directional pinning is likely to exist. 
This is a direct test of the presence of directional pinning in the flow+Campbell regime.

A complete picture is obtained once the intrinsic $\intr\rho_{\ff,11}$ is obtained.
If $\intr\rho_{\ff,11}$ is found to satisfy the scaling, and in the same time $r$ or, equivalently $\intr\rho_{C,11}$, is not, this result would constitute evidence for
the presence of directional pinning contributions such as extended defects. 
This type of analysis was performed in references \cite{Pompeo2012a, Pompeo2012b, Pompeo2013}, where it enabled the accurate extraction of the intrinsic anisotropy of BaZrO$_3$ added YBa$_2$Cu$_3$O$_{7-\delta}$ thin films, the unambiguous identification in the angular dependent pinning constant of extended pinning acting effectively at microwave regimes, and the identification of a Mott-insulator effect through a comparative study with d.c. $J_c$ measurements.

Thus, the application of the results of this paper to angular measurements provides useful tools to ascertain several issues linked to the anisotropy of superconducting materials: from the identification of directional pinning, to the determination of the mass--anisotropy $\gamma$ factor and of geometry--independent quantities (such as $r$ or $\omega_p$).

\section{Summary}
\label{sec:summary}

In this paper we have developed a tensor model for the full a.c. vortex motion complex resistivity, which includes free flux--flow, elastic recall from pinning centers, and flux creep. The model takes into account the effect of a magnetic field applied to an arbitrary angle with respect to the crystal directions of a uniaxially anisotropic superconductor and with respect to the applied current density. 
We have shown that intrinsic vortex parameters (``material properties'') and directly measured vortex parameters may differ due to the geometry of the experiment. We have developed a procedure to extract the intrinsic properties from experiments, and we have shown the main differences in several commonly used experimental geometries. Finally, we have shown that a combination of angular measurements and an analysis in terms of the present model joint to the angular scaling prescription can give a large amount of relevant information, including direct evidence of directional pinning.
Future extensions of this work are (i) the derivation of the full surface impedance expression in the mixed state including both the anisotropic complex vortex resistivity tensor here presented and the contribution of the quasi-particles and Cooper pairs; and 
(ii) the determination of the pinning constant tensor when extended pinning centers are present, a matter of large interest within the present research focused on the addition of artificial pinning centers in High-$T_c$ superconductors.

\section*{Acknowledgments}
This work has been carried out within the framework of the EUROfusion Consortium and has received funding from the Euratom research and training programme 2014-2018 under grant agreement No 633053. The views and opinions expressed herein do not necessarily reflect those of the European Commission.
 
\begin{appendices}
\section{Straight planar currents}
\label{sec:rhovm_examples1}

A straight a.c. current can be applied to flat thin films using, e.g.,  resonators with rectangular geometries \cite{Torokhtii2014,Pompeo2017},
or placing a small portion of the sample in the appropriate region of a resonant cavity.
The vortex resistivity measured along the $\vec{J}\parallel\uvec{x}$ direction can be computed as $\left(\tens{\rho_v}\uvec{x}\right)\cdot\uvec{x}$ using the tensor \eref{eq:rho_v}:
\begin{subequations}
\label{eq:rhoeff_exp4}
\begin{align}
\label{eq:rhoeff_exp4a}
\rho_{v}^{(x)}(B,\theta,\phi)&=\intr\rho_{v,11}(B,\theta)f_{L\phi}(\theta,\phi)\\
\label{eq:rhoeff_exp4b}
f_{L\phi}(\theta,\phi)&=\frac{\gamma^{-2}\sin^2\theta\sin^2\phi+\cos^2\theta}{\gamma^{-2}\sin^2\theta+\cos^2\theta} 
\end{align}
\end{subequations}

\begin{figure}[ht]
\centerline{(a) \includegraphics[width=0.3\columnwidth]{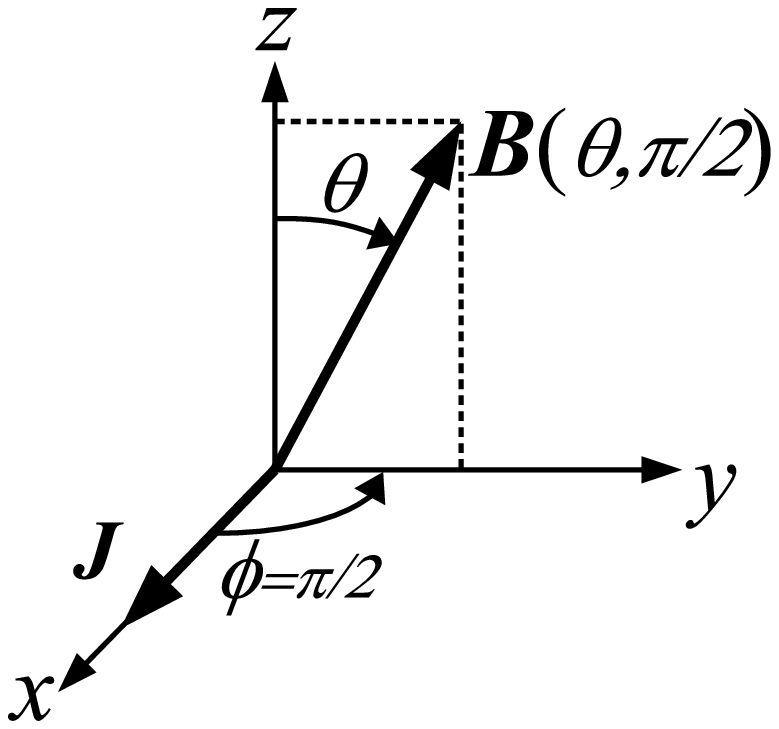} (b) \includegraphics[width=0.3\columnwidth]{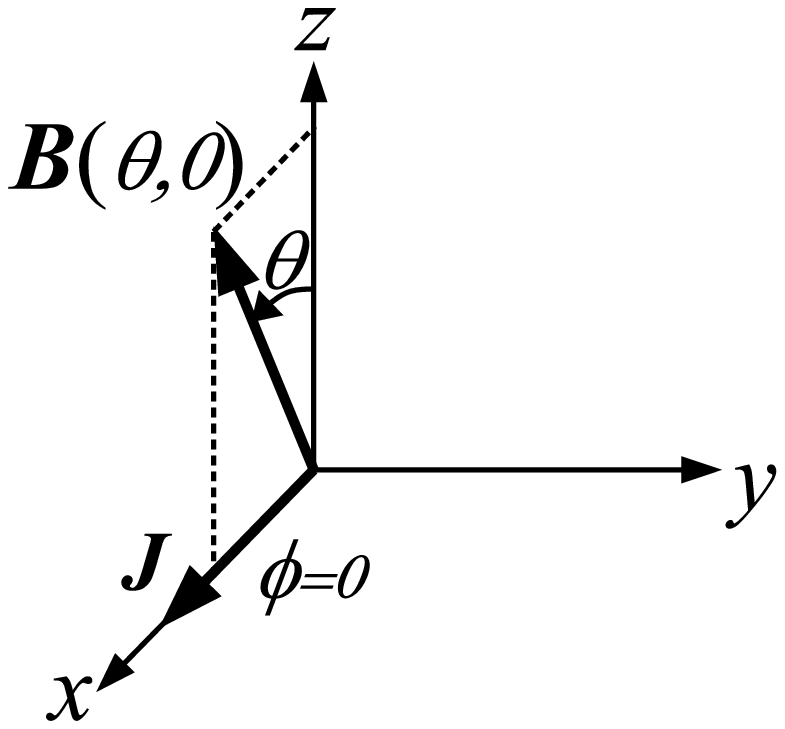}}
  \caption{
  Most common experimental configurations with straight currents. (a): ``maximum Lorentz force configuration''; the applied field is always perpendicular to the current, which flow in the isotropic plane. (b): ``variable Lorentz force configuration''; the projection of the applied field to the isotropic plane is always along the current.
}
\label{fig:straight}
\end{figure}

In practical cases, two particular configurations deserve some discussions. First, in the so--called ``maximum Lorentz force configuration'', where the applied field is always perpendicular to the current ($\phi=\pi/2$, see Fig. \ref{fig:straight}a), one has $f_{L\phi}(\theta,\phi)=1$ and the rather expected result $\rho_{v}^{(x)}(B,\theta,\phi)=\intr\rho_{v,11}(B,\theta)$. In this case, one has direct, experimental access to the intrinsic vortex resistivity $\intr\rho_{v,11}$.
One would conclude that this configuration is the best geometry to directly access the intrinsic vortex resistivity.

A second geometry with $\phi = 0$ has been used both in d.c. \cite{Iye1989, Jakob1993} and at microwave frequencies \cite{Lofland1995b}.
Here, the plane determined by the tilted field and $\vec{J}$ (flowing in the $ab$--plane) is perpendicular to the $ab$ plane, see Fig. \ref{fig:straight}b. In this case one has clearly an angle--dependent Lorentz--force contribution, and $f_{L}(\theta)=1/({\gamma^{-2}\tan^2\theta+1})$. For $\theta=\pi/2$ (no Lorentz force) one has $f_L=0$, recovering the known results for scalar models. However, when the field is at an arbitrary angle, the widespread habit is to consider the variable Lorentz force as an additional $\cos^2\theta$ term \cite{Lofland1996}. This is true for isotropic materials ($\gamma=1$), but it is not correct for anisotropic superconductors. The discrepancy has its roots in the nontrivial motion of vortices in an anisotropic medium: there, vortex motion is not in general perpendicular to the exerted Lorentz force, and the induced electric field is not parallel to the applied current.
Of course, the extent to which $f_L(\theta)\neq\cos^2\theta$ depends on $\gamma$, and the influence on the measurements has to be evaluated depending on the material. Nevertheless, it is a factor to be considered in the analysis of the experiments.

\section{Useful expressions}
\label{sec:useful}
In the analysis of the data one usually looks for the information contained in the ``intrinsic'' quantities $\intr\rho_{ff}, \intr\eta, \intr{k}_p, \omega_p$, but experiments yield $\rho_v$. In addition, anisotropy and Lorentz force contributions lead to averaged quantities over the current paths, whence effective quantities.
In Table \ref{tab:useful} we collect the 
expressions that should be used in
 the analysis of experimental data
to extract the intrinsic quantities from experimental values. We focussed
on the GR model, hence neglecting creep.
Rows 1 and 2 contain the first step of the analysis. Row 3 is the definition (given for completeness) of the effective (current-path averaged) flux-flow resistivity. Rows 4, 5, and 10 (6, 7, and 11) give the further steps for rotational (linear) symmetric currents. Row 8, 12, 13 represent the expectations of the BGL scaling, with the angular scaling factor given in Row 9. In order to check for, e.g., directional pinning, one should look for discrepancies in the angular dependencies of $\intr{k}_{p,11}$ as given by Row 12 (theory) and the experimental values given by Row 10 or 11.

\begin{table}[h]
\begin{tabular} {cll}
\hline
& Expression & ref.\\
\hline
\\[-1em]
1 & $\rho_{\ff,\eff}  =\Re(\rho_{v,\eff}) \left[1+\left(\frac{\Im(\rho_{v,\eff})}{\Re(\rho_{v,\eff})}\right)^2\right]$ & \eqref{eq:rhoGR2a} \\
2 & $k_{p,\eff} = \omega B \Phi_0 \frac{\Im(\rho_{v,\eff})}{\Re^2(\rho_{v,\eff})+\Im^2(\rho_{v,\eff})}$ & \eqref{eq:rhoGR2c} \\
3 & $\rho_{\ff,\eff}(B,\theta, \phi)=\frac{1}{S}\int_0^{2\pi}\left(\tens{\rho_\ff}(B,\theta,\phi)\uvec{J}(\vec{r})\right)\cdot\uvec{J}(\vec{r}) dS$ & \cite{Collin1992}\\
4 & $\rho_{\ff,\eff}(B,\theta) =\intr\rho_{\ff,11}(B,\theta) f_{L}(\theta)$ & \eqref{eq:rhoGR3a} \\
5 & $f_{L}(\theta)=\frac{\frac{1}{2}\gamma^{-2}\sin^2\theta+\cos^2\theta}{\gamma^{-2}\sin^2\theta+\cos^2\theta}$ & \eqref{eq:rhoeff_exp53b}\\
6 & $\rho_{\ff,\eff}^{(x)}(B,\theta,\phi)=\intr\rho_{\ff,11}(B,\theta)f_{L\phi}(\theta,\phi)$ & \eqref{eq:rhoeff_exp4a}\\
7 & $f_{L\phi}(\theta,\phi)=\frac{\gamma^{-2}\sin^2\theta\sin^2\phi+\cos^2\theta}{\gamma^{-2}\sin^2\theta+\cos^2\theta}$ & \eqref{eq:rhoeff_exp4b}\\
8 & $\intr\rho_{\ff,11}(B,\theta)={\intr\rho}_{\ff,11}(B\epsilon(\theta),0)$ & \eqref{eq:etatensor_scalinga}\\
9 & $\epsilon(\theta)=(\cos^2\theta+\gamma^{-2}\sin^2\theta)^{1/2}$ & \eqref{eq:epsilon} \\
10 & $k_{p,\eff}(B,\theta)=\frac{\intr k_{p,11}(B,\theta)}{f_{L}(\theta)}$ & \eqref{eq:rhoGR3c}\\
11 & $k_{p,\eff}(B,\theta,\phi)=\frac{\intr k_{p,11}(B,\theta)}{f_{L\phi}(\theta,\phi)}$ & \\
12 &  $\intr k_{p,11}(B,\theta)=\frac{\intr k_{p,11}(B\epsilon(\theta),0)}{\epsilon(\theta)}$ & \eqref{eq:kptensor_Bdep} \\ 
13 & $\frac{{\intr k}_{p,ii}(B,\theta)}{\intr\eta_{ii}(B,\theta)}=\omega_p(B,\theta)=\omega_p(B\epsilon(\theta),0)$ & \eqref{eq:omegap} \\
\hline
\vspace{1mm}
\end{tabular}
\caption{
Selected useful expressions, with the references to their actual appearance in the body of the manuscript. }
\label{tab:useful}
\end{table}

\end{appendices}

\end{document}